\documentclass[journal]{IEEEtran}

\usepackage[cmex10]{amsmath}
\usepackage{amssymb}
\usepackage[mathscr]{eucal}
\usepackage{cite}
\usepackage{amsbsy}
\usepackage{esint}
\usepackage{tikz}
\usepackage{graphicx}
\usepackage{multirow}

\newtheorem{observation}{Observation}[section]

                    {$\blacksquare$\vspace*{7pt}} 

\def\R{{\mathbb R}}

\def\su{\mathrm{supp}}

\def\CT{\mathrm{CT}}

\def\su{\mathrm{supp}}
\def\Om{\Omega}

\def\f{\frac}

\def\bi{{\mathbf i}}

\def\x{\boldsymbol{x}}

\def\mR{{\mathcal R}}

\def\Ef{{\mbox{{\footnotesize $E$}}}}

\def\bi{\begin{itemize}} \def\ei{\end{itemize}}
\def\be{\begin{eqnarray*}}
\def\ee{\end{eqnarray*}}

\def\etal{{\it et al }}

\def\0{{\mathbf 0}}

\newcommand{\beq}{\begin{equation}}
\newcommand{\eeq}{\end{equation}}

\def\eref#1{(\ref{#1})}

\def\XXint#1#2#3{{\setbox0=\hbox{$#1{#2#3}{\int}$ }
\vcenter{\hbox{$#2#3$ }}\kern-.55\wd0}}

\ifCLASSINFOpdf

\else

\fi

\usepackage{algorithmic}

\usepackage{array}

\usepackage{fixltx2e}

\usepackage{stfloats}

\usepackage{url}

\begin{document}

\title{Machine-learning-based nonlinear decomposition of CT images for metal artifact reduction}


\author{\IEEEauthorblockN{Hyung Suk Park\IEEEauthorrefmark{1}, Sung Min Lee\IEEEauthorrefmark{2}, Hwa Pyung Kim\IEEEauthorrefmark{2}, and Jin Keun Seo\IEEEauthorrefmark{2}}\\
\IEEEauthorblockA{\IEEEauthorrefmark{1} Division of Strategic Research, National Institute for Mathematical Sciences, Daejeon, 34047, South Korea}\\
\IEEEauthorblockA{\IEEEauthorrefmark{2} Department of Computational Science and Engineering, Yonsei University, Seoul, 120-749, South Korea}
\thanks{Manuscript received XXX; revised  XXX.
Corresponding author: S. M. Lee (email: sungminlee@yonsei.ac.kr).}}

\markboth{}%
{Park \MakeLowercase{\textit{et al.}}: }

\IEEEtitleabstractindextext{%
\begin{abstract}
Computed tomography (CT) images containing metallic objects commonly show severe streaking and shadow artifacts. Metal artifacts are caused by nonlinear beam-hardening effects combined with other factors such as scatter and Poisson noise. In this paper, we propose an implant-specific method that extracts beam-hardening artifacts from CT images without affecting the background image. We found that in cases where metal is inserted in the water (or tissue), the generated beam-hardening artifacts can be approximately extracted by subtracting artifacts generated exclusively by metals. We used a deep learning technique to train nonlinear representations of beam-hardening artifacts arising from metals, which appear as shadows and streaking artifacts. The proposed network is not designed to identify ground-truth CT images (i.e., the CT image before its corruption by metal artifacts). Consequently, these images are not required for training. The proposed method was tested on a dataset consisting of real CT scans of pelvises containing simulated hip prostheses. The results demonstrate that the proposed deep learning method successfully extracts both shadowing and streaking artifacts.

\end{abstract}

}

\maketitle

\IEEEdisplaynontitleabstractindextext

\IEEEpeerreviewmaketitle

\section{Introduction}
X-ray computed tomography (CT) provides tomographic images of the human body by assigning an X-ray attenuation coefficient to each pixel, using sinogram data (i.e., X-ray data collected at all angles around the object). The reconstruction methods used for CT images are based on a linear assumption of the sinogram data: namely, the Radon transform of an image. However, this assumption of linearity is violated in the presence of highly attenuating materials such as metal implants (e.g., hip replacements, dental fillings, surgical clips, and pacemaker wires) in the CT scan field. These discrepancies in the linear model are mainly caused by beam-hardening effects, which are associated with the polychromatic nature of the X-ray beam and the energy-dependent variations in the attenuation coefficients of highly attenuating materials. The inconsistent sinogram data attributable to these beam-hardening effects do not match the Radon transform of any attenuation coefficient distribution. Consequently, metallic artifacts seriously degrade the CT images of patients with metal implants.

Because metallic implants are increasingly popular, there is a growing demand for metal artifact reduction (MAR) in dentistry and medicine. Despite numerous efforts to develop MAR methods in the last four decades, it remains a very challenging problem because of the difficulties presented by the nonlinear effects that arise from the geometry of high-attenuation materials, polychromatic X-ray beams, and related artifacts, together with Compton scattering, photon noise, and so on. Existing MAR methods include dual-energy CT\cite{Alvarez1976}, statistical iterative-reconstruction methods\cite{DeMan2001,Elbakri2002,Menvielle2005,OSullivan2007,Wang1996}, data completion/inpainting-based methods\cite{Abdoli2010,Bazalova2007,Kalender1987,Lewitt1978,Meyer2010,Park2013,Roeske2003}, and hybrid methods\cite{Lemmens2009}. However, several current methods produce new artifacts that are not present before the application of the method\cite{Meyer2010, Muller2009}.

In this paper, we propose a deep learning method for implant-specific MAR that extracts beam-hardening artifacts from CT images. In cases where metal inserts are placed inside of tissue (or water), we found that metal-induced beam-hardening artifacts can be approximately extracted by subtracting artifacts generated exclusively by metal inserts. With this observation, the proposed MAR method takes advantage of training with metal-only images to extract features of the beam-hardening artifacts.

Because beam-hardening artifacts depend on the metal geometry, energy dependency of the attenuation coefficient, and the spectrum of the incident X-ray beam, the number of variations is too great to allow the features to be learned. This obstacle is dealt with by incorporating prior knowledge of the geometry of the metallic objects. Indeed, the geometry and material information of implants such as hip replacements and endovascular stents are usually known.  With this knowledge of the scanned metal, we adopt a multi-scale convolutional network, called U-net \cite{Ronneberger2015}, to train the model using metal-only images.

In the presence of metallic objects, no ground-truth CT images (i.e.,
CT images before corruption by metal artifacts) are available. In such cases, they cannot be used as label data for learning. The proposed method does not require ground-truth image, insofar as it learns artifacts generated exclusively by metal, rather than artifact-free CT images.

We explore the feasibility of applying a deep learning approach to MAR using a dataset consisting of real CT scan of pelvises containing two simulated hip prostheses. The results demonstrate that the proposed deep learning method successfully extracts both shadowing and streaking artifacts.

\section{Method}\label{analysis_sec}

Considering a two-dimensional parallel beam, we define X-ray data for a polychromatic X-ray along the rotation direction $(\cos\varphi,\sin\varphi),~\varphi\in[0,2\pi)$ as
\begin{align}\label{P_f}
P_{f}(\varphi,s)=-\ln\left(\int_{0}^{\overline{E}}\eta(E)e^{-\mR f_E(\varphi,s)}dE\right),
\end{align}
where $f_{\Ef}(\x)$ denotes the distribution of the attenuation coefficient at  position $\x=(x_1,x_2)$ and at energy level $E$,
$\eta(\Ef)$ represents the fractional energy at photon energy $E$ in the spectrum of the X-ray source \cite{Herman1983,Poludniowski2009}, that is, $\int_0^{\overline{E}} \eta(E) dE=1$, and $\mR$ denotes the Radon transform \cite{Radon2005}.  The most widely used reconstruction method for CT is the filtered backprojection algorithm (FBP):
\begin{align}\label{fbp}
f_{\CT}(\x):=\mR^{-1}P_f(\x),
\end{align}
where $\mR^{-1}$ denotes the filtered backprojection operator \cite{Bracewell1967}.

\subsection{Metal-induced beam-hardening artifact extraction}
\begin{figure}[ht]
\centering
\includegraphics[width=0.5\textwidth]{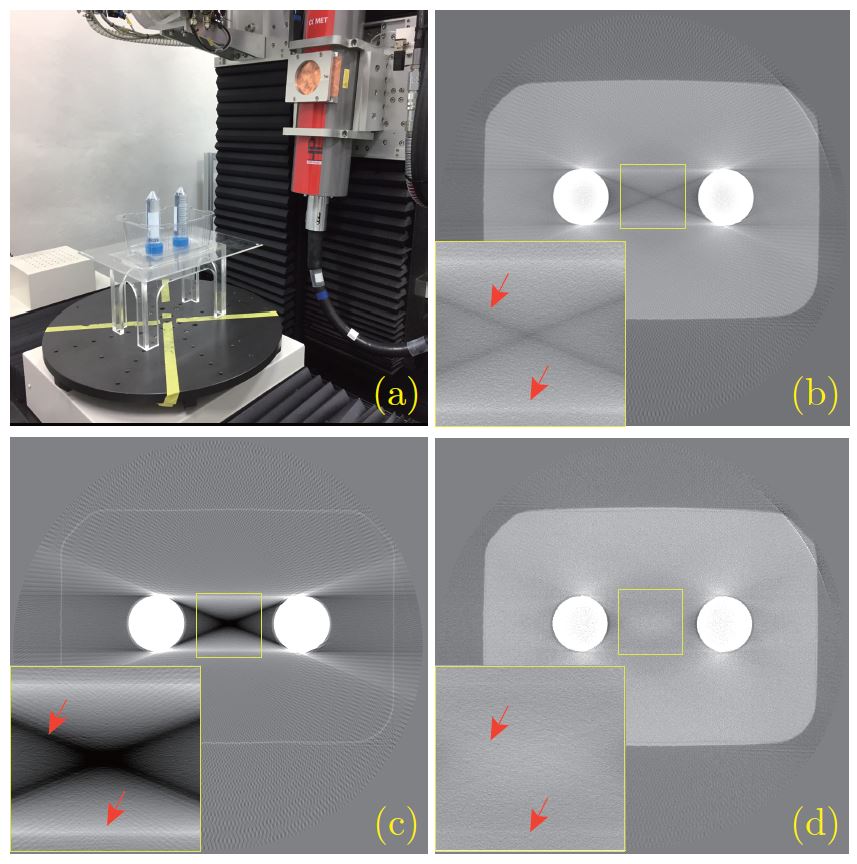}
\caption{Illustration of Observation \ref{prop1}: (a) water phantom consisting of water and two cylinders filled with highly attenuating fluid; (b) reconstructed image $f_\CT$; (c) metal-induced artifact  $f^m$; and (d)  $f_\CT-\gamma f^m$ with $\gamma=0.4$. (C=-1000 HU/W=4000 HU for CT images.)}
\label{fig-intro}
\end{figure}

In this section, we propose a method of extracting beam-hardening artifacts arising from metallic objects from the measured $f_\CT$. Let the cross-sectional slice to be scanned occupy a domain $\Om$. Assume that $f_E$ can be expressed as
\begin{align}
  f_E(\x)=\mu_t(E)\chi_{_{D_t}}(\x)+\mu_m(E)\chi_{_{D_m}}(\x),~~ \x\in\Om,
\end{align}
where $\mu_t$ and $\mu_m$ denote the linear attenuation coefficients of tissue and the metal in the domain $D_t$, $D_m\subset\Om$, respectively. Here, $\chi_D$ denotes the characteristic function, which is 1 in the region $D$ and 0, otherwise.  Let $P_{f^m},P_{f^t}$ be the projection data for metal and tissue given by
  \begin{align}\label{pfm}
  P_{f^m}(\varphi,s)=-\ln\left(\int_{0}^{\overline{E}}\eta(E)e^{-\mu_m(E)\mR \chi_{_{D_m}}(\varphi,s)}dE\right),
  \end{align}
  and
  \begin{align}
  P_{f^t}(\varphi,s)=-\ln\left(\int_{0}^{\overline{E}}\eta(E)e^{-\mu_t(E)\mR \chi_{_{D_t}}(\varphi,s)}dE\right),
  \end{align}
  respectively. Let $f^m$ and $f^t$ be the CT images for $P_{f^m}$ and $P_{f^t}$, respectively. The following observation expresses the extraction of metal artifacts ($f^m$) from $f_\CT$  (see also Fig.\ref{fig-intro}):

\begin{observation}\label{prop1}
  Let $D_m$ be a homogeneous metal region inserted in the tissue. Then, the $f_\CT$ in \eref{fbp} can be  approximately decomposed into
  \begin{align}\label{ob-eq}
    f_\CT(\x)\approx\tau\chi_{_{D_m}}(\x) + f^t(\x) + \gamma f^m(\x),
  \end{align}
  where $\tau,\gamma$ are the constants depending on the distribution of $\mu_m$ and $\mu_t$. In the absence of tissue, that is, $f^t(\x)=0$, for all $\x\in\Om$, $\tau=0$ and $\gamma=1$ .
\end{observation}

Figure \ref{fig-intro} illustrates Observation \ref{prop1}. Fig. \ref{fig-intro}(a) shows a phantom consisting of water and two cylinders filled with highly attenuating fluid. The fluid consists of iodinated contrast media and saline. Fig. \ref{fig-intro}(b) shows the corresponding $f_\CT$ obscured by artifacts arising from this fluid. Figs. \ref{fig-intro}(c) and (d) show $f^m$ and $f_\CT-\gamma f^m$ with $\gamma = 0.4$, respectively.  Fig. \ref{fig-intro}(d) shows that the beam-hardening artifacts are almost completely removed from $f_\CT$. In addition, because $f^m$ is subtracted from $f_\CT$, streaking artifacts due to under-sampling are significantly reduced. In this experiment, CT images were acquired from an industrial CT scanner (DUKIN, Korea) with an X-ray tube voltage of 100 kV and a tube current of 12 mA.

Now, let us discuss the meaning of Observation \ref{prop1}. Equation \eref{ob-eq} implies that for all X-rays passing through metallic objects, two $\mu_m$ at mean beam energies in the presence and absence of tissue have a nearly-linear relationship.  More specifically, the projection data corresponding to \eref{ob-eq} is given by
\begin{align}\label{ob-eq1}
P_f(\varphi,s)\approx\tau\mR\chi_{_{D_m}}(\varphi,s) + P_{f^t}(\varphi,s) + \gamma P_{f^m}(\varphi,s).
\end{align}
Because $\int_0^{\overline{E}} \eta(E) dE=1$, it follows from the mean value theorem for integration that for each $(\varphi,s)\in [0,2\pi)\times\R$, there exists $E^c_{\varphi,s}, E^m_{\varphi,s}\in [0,\overline{E}]$, such that
\begin{align}\label{mvt-ct}
& \int_{0}^{\overline{E}}\eta(E)e^{-\mu_m(E)\mR \chi_{_{D_m}}(\varphi,s)-\mu_t(E)\mR \chi_{_{D_t}}(\varphi,s)}dE\nonumber\\
 &\quad\quad= e^{-\mu_m(E^c_{\varphi,s})\mR \chi_{_{D_m}}(\varphi,s)}\int_{0}^{\overline{E}}\eta(E)e^{-\mu_t(E)\mR \chi_{_{D_t}}(\varphi,s)}dE
\end{align}
and
\begin{align}\label{mvt-m}
\int_{0}^{\overline{E}}\eta(E)e^{-\mu_m(E)\mR \chi_{_{D_m}}(\varphi,s)}dE= e^{-\mu_m(E^m_{\varphi,s})\mR \chi_{_{D_m}}(\varphi,s)}.
\end{align}

It follows from \eref{mvt-ct} and \eref{mvt-m} that \eref{ob-eq1} can be expressed as
\begin{align}\label{ob-eq2}
&\mu_m(E^c_{\varphi,s})\mR \chi_{_{D_m}}(\varphi,s)\\
&\qquad \approx \tau\mR\chi_{_{D_m}}(\varphi,s)+\gamma~ \mu_m(E^m_{\varphi,s})\mR \chi_{_{D_m}}(\varphi,s).
\end{align}
This leads to
\begin{align}\label{ob-eq3}
\mu_m(E^c_{\varphi,s})\approx\gamma \mu_m(E^m_{\varphi,s})+\tau, \mbox{for }
(\varphi,s)\in\su(\mR\chi_{_{D_m}})
\end{align}
where $\su(\mR\chi_{_{D_m}})$ denotes the support of $\mR\chi_{_{D_m}}$,  defined as $\{(\varphi,s)\in [0,2\pi)\times\R:\mR\chi_{_{D_m}}(\varphi,s)\neq0\}$.

\begin{figure}[ht]
\centering
\includegraphics[width=0.5\textwidth]{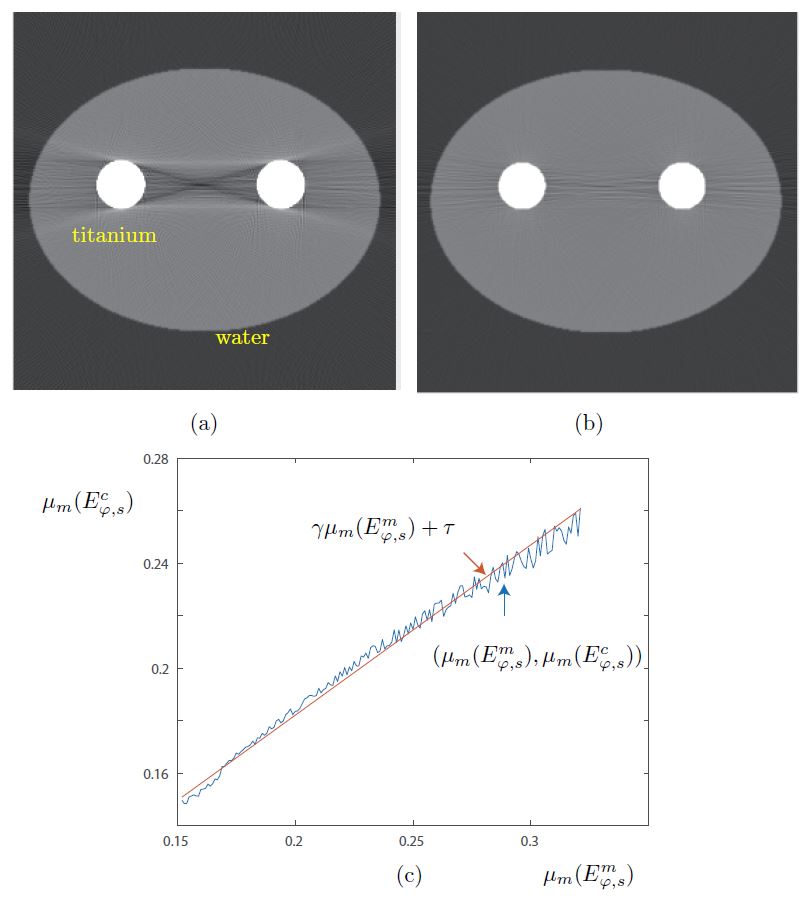}
\caption{(a) $f_\CT$ for water phantom containing two metallic inserts (titanium); (b) $f_\CT-\gamma f^m$ with $\gamma =0.65$ and $\tau =0.051$; and (c) $\mu_m(E^c_{\varphi,s})$-$\mu_m(E^m_{\varphi,s})$ plot for the phantom in (a)}
\label{fig-ob1}
\end{figure}

\begin{figure*}[ht]
\centering
\includegraphics[width=1\textwidth]{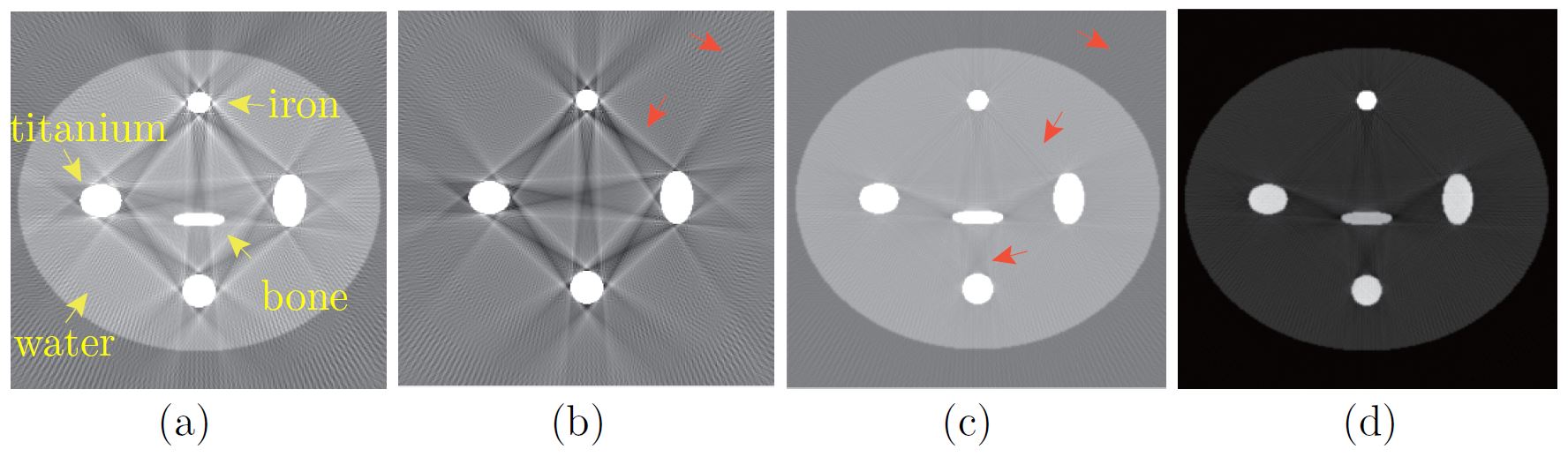}
\caption{Results from extracting metal-induced beam-hardening artifacts. Figures (a) and (b) show $f_\CT$ for a water phantom containing four metallic inserts (three titanium and one iron) and $f^m$, respectively. Figures (c) and (d) show $f_\CT-\gamma f^m$ with $\gamma =0.74$ at different window levels. (C=-500 HU/W=3000 HU for (a), (b) and (c), and C=2500 HU/W=6000 HU for (d).)}
\label{fig-val}
\end{figure*}

We show convincing evidence from numerical simulation that the relation in Equation \eref{ob-eq3} is valid. Fig. \ref{fig-ob1} (a) shows the CT image with a water phantom containing two metallic objects (viz., titanium). Fig. \ref{fig-ob1} (c) shows the plot of $\mu_m(E^c_{\varphi,s})$ along the $y$-axis versus the corresponding $\mu_m(E^m_{\varphi,s})$ along the $x$-axis. As shown in Fig. \ref{fig-ob1} (c), these two variables satisfy a nearly linear relationship for all $(\varphi,s)\in [0,2\pi)\times\R$. This relation can be approximated by the line $\gamma \mu_m(E^m_{\varphi,s})+\tau$ with $\gamma =0.65$ and $\tau =0.051$, and the corresponding reconstructed image $f_\CT-\gamma f^m$ almost completely removes the metal-induced beam-hardening artifacts, as shown in Fig. \ref{fig-ob1} (b). In this simulation, $\mu_m(E^c_{\varphi,s})$ and $\mu_m(E^m_{\varphi,s})$ were computed from the relation of \eref{mvt-ct} and \eref{mvt-m}, which, for  $(\varphi,s)\in\su(\mR\chi_{_{D_m}})$, is given by
\begin{align}
\mu_m(E^c_{\varphi,s}) = \f{P_f(\varphi,s)-P_{f^t}(\varphi,s)}{\mR\chi_{_{D_m}}(\varphi,s)},
\end{align}
and
\begin{align}
\mu_m(E^m_{\varphi,s}) = \f{P_{f^m}(\varphi,s)}{\mR\chi_{_{D_m}}(\varphi,s)},
\end{align}
respectively. Note that there remain streaking artifacts along the X-ray line passing through both two titanium inserts. This is because of photon starvation \cite{Barrett2004}. We placed three titanium inserts, one iron insert, and one bone insert, all of different shapes and sizes, inside of water phantom, as shown in Fig. \ref{fig-val} (a). The relation  \eref{ob-eq} in Observation \ref{prop1} seems to be valid for more complex metallic inserts; beam-hardening artifacts are effectively extracted by subtracting $f^m$ from $f_\CT$. However, the beam-hardening artifacts in $f^m$ cannot deal with artifacts between metallic inserts and bone. These still remain in $f_\CT-\gamma f^m$, as shown in Fig. \ref{fig-val} (c). More careful analysis is needed to deal with such artifacts.

In order to extract $f^m$ from the measured $f_\CT$, $f^m$ can be computed numerically from \eref{pfm}. This requires the domain occupying the metallic region $D_m$, the X-ray spectrum $\eta$, and the linear attenuation coefficient of the metal object $\mu_m$ in $[0, \overline{E}]$. However, the information regarding $\eta$ and $\mu_m$ might not be (accurately) provided for commercial CT scanners and scanned metallic inserts. Moreover, discretization methods for integrating \eref{pfm} can lead to discretization errors.  Alternatively, in the following we apply a deep convolutional neural network (CNN) technique to estimate $f^m$ directly.


\subsection{Deep learning method for generating metal beam-hardening artifacts}

To learn the features of metal-induced beam-hardening artifacts, we need to consider a suitable dataset and CNN. Because $f^m$ depends on the metal geometry, the energy distribution of the X-ray beam, and the incident X-ray spectrum, the number of variations is too great to allow the features to be learned. Hence, it is crucial to choose an input image that takes into account such information.

Recently, Park \etal showed that $f^m$ can be approximated by the nonlinear function $g^m$, given as \cite{Park2016a}
\begin{align}\label{gm}
g^m(\x) = -\mR^{-1}\left[\ln\left(\f{\sinh(\lambda\mR\chi_{_{D_m}})}{\lambda\mR\chi_{_{D_m}}}\right)\right](\x) + c\chi_{_{D_m}}(\x),
\end{align}
where $\lambda$ is the parameter depending on $\mu_m$ and $\eta$, and $c$ is some constant. In other words, the measured $f^m$ from a CT scan can be approximated without the use of any prior knowledge of $\mu_m$ and $\eta$.

In our deep learning network, the goal is to learn the map from the formula-based artifact $g^m$ to the corresponding true artifact $f^m$. We used a labeled dataset $\{(g^m_i, f^m_i): i=1,2,\cdots\}$ to find the map, $\Upsilon:g^m\mapsto f^m$, that minimizes the $L^2$ error $Err(\Upsilon)$, given by
\begin{align}\label{error}
Err(\Upsilon)=\|f^m-\Upsilon(g^m)\|_{L^2(\R^2)}^2.
\end{align}
We apply a multi-scale convolutional network, called U-net, to estimate $\Upsilon$. Unlike traditional deep CNNs \cite{LeCun1989}, this network does not have a fully connected layer. Hence, U-net can learn large-scale input images (i.e., $g^m$). In addition, simulation results show that this multi-scale network can efficiently learn global features such as streaking \cite{Gu2017,Han2016}.

\begin{figure*}[ht]
\centering
\includegraphics[width=1\textwidth]{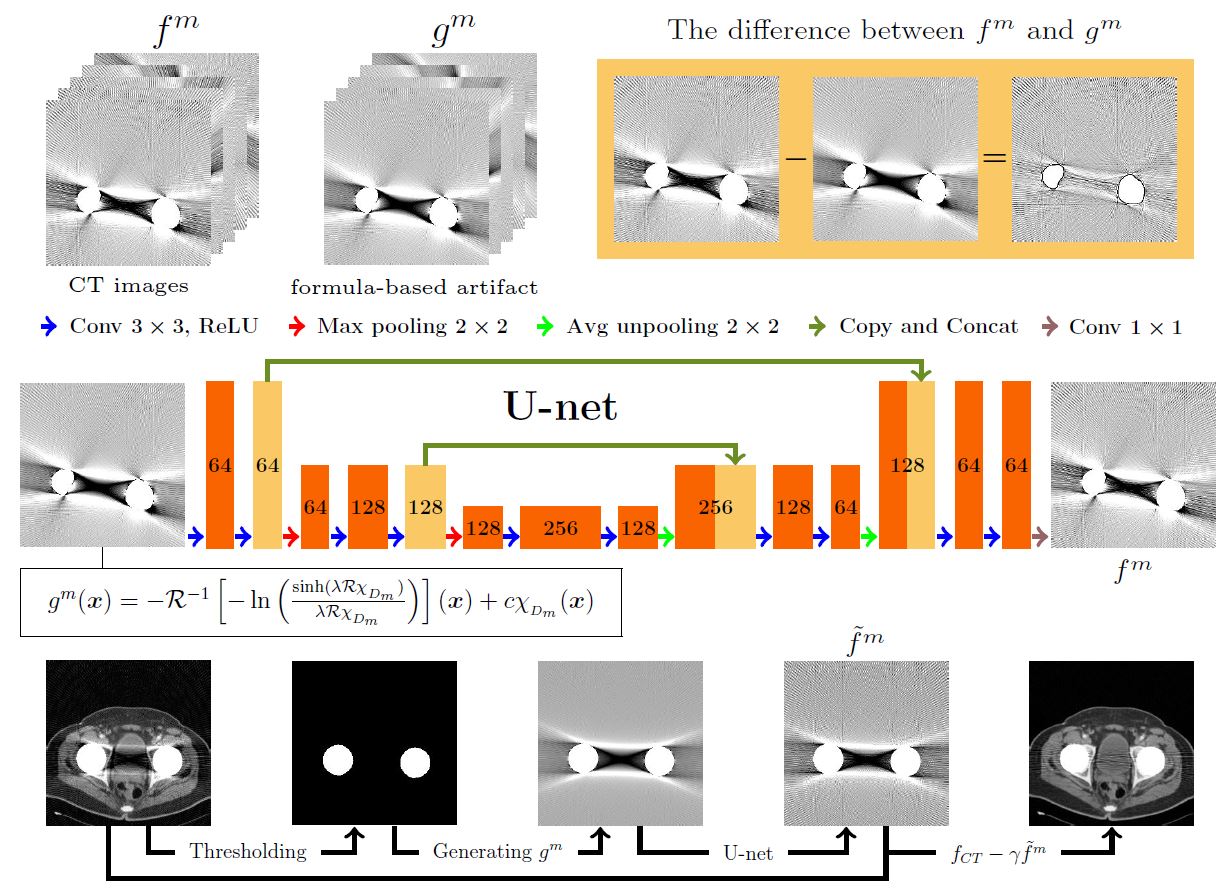}
\caption{Illustration of the multi-scale convolutional network for training $f^m$ from $g^m$, along with the algorithm for the proposed method.}
\label{unet}
\end{figure*}

The architecture of the multi-scale convolutional network is shown in Fig. \ref{unet}. The architecture consists of a contracting path and an expansive path. Each step of the contracting path contains two convolutions with a $3\times3$ window, each followed by a rectified linear unit (ReLU), along with $2\times 2$ max pooling with strides of two in each direction of the domain. In each step of the expansive path, the average unpooling is used instead of max pooling. Then, it is concatenated with the features in the contracting path during the same step. Note the every convolution in our network is performed with zero-padding in order to match the size of the input and label images.

\subsection{Algorithm}
The proposed method is based on the following steps (see also Fig. \ref{unet}):
\begin{itemize}
  \item[(i)] From the measured $P_f$, reconstruct the uncorrected $f_\CT$ using the FBP algorithm.
  \item[(ii)] Generate $g^m$ from the formula \eref{gm}, and then generate the output $\tilde{f}^m$ from the trained network.
  \item[(iii)] Compute corrected CT image $f_\CT-\gamma \tilde{f}^m$.
\end{itemize}
In Step (ii), $g^m$ in \eref{gm} is determined by the region $D_m$ and parameters $\lambda$ and $c$, which are obtained by the following procedure. First, the $D_m$ is extracted with simple thresholding from $f_\CT$, and $\lambda$ is determined as in \cite{Park2016a}. From the relation of \eref{gm} and $f^m\approx g^m$, $c$ is computed as follows:
\begin{align}
  c = \f{1}{|\chi_{_{D_m}}|}\int_{D_m}f^m(\x)+\mR^{-1}\left[\ln\left(\f{\sinh(\lambda\mR\chi_{_{D_m}})}{\lambda\mR\chi_{_{D_m}}}\right)\right](\x)~d\x.
\end{align}
Here, $c$ is computed as the average value over $D_m$. In Step (iii), we determine $\gamma$ in such a way that any streaking artifacts arising from metals are alleviated in $f_\CT-\gamma \tilde{f}^m$. More specifically, we find the $\gamma$ minimizing the function $\Phi$ given by
\begin{align}\label{ob-fn}
  \Phi(\gamma) = \int_{\Om\setminus D_m} |\nabla(f_\CT(\x)-\gamma \tilde{f}^m(
  \x))|^2 d\x.
\end{align}
Because $\Phi(\gamma)$ is one-dimensional function with respect to $\gamma$, one can easily find minimizer $\gamma$ by computing $\Phi(\gamma)$ over the suitable interval.

Note that the error $g^m-f^m$ comes from model's assumption used to derived the formula $g^m$, for example, from the use of a linear approximation for nonlinear variation in the attenuation coefficient \cite{Park2016a} (see Fig. \ref{unet}).

\section{Results}
\subsection{Dataset}
\begin{figure*}[ht]
\centering
\includegraphics[width=.8\textwidth]{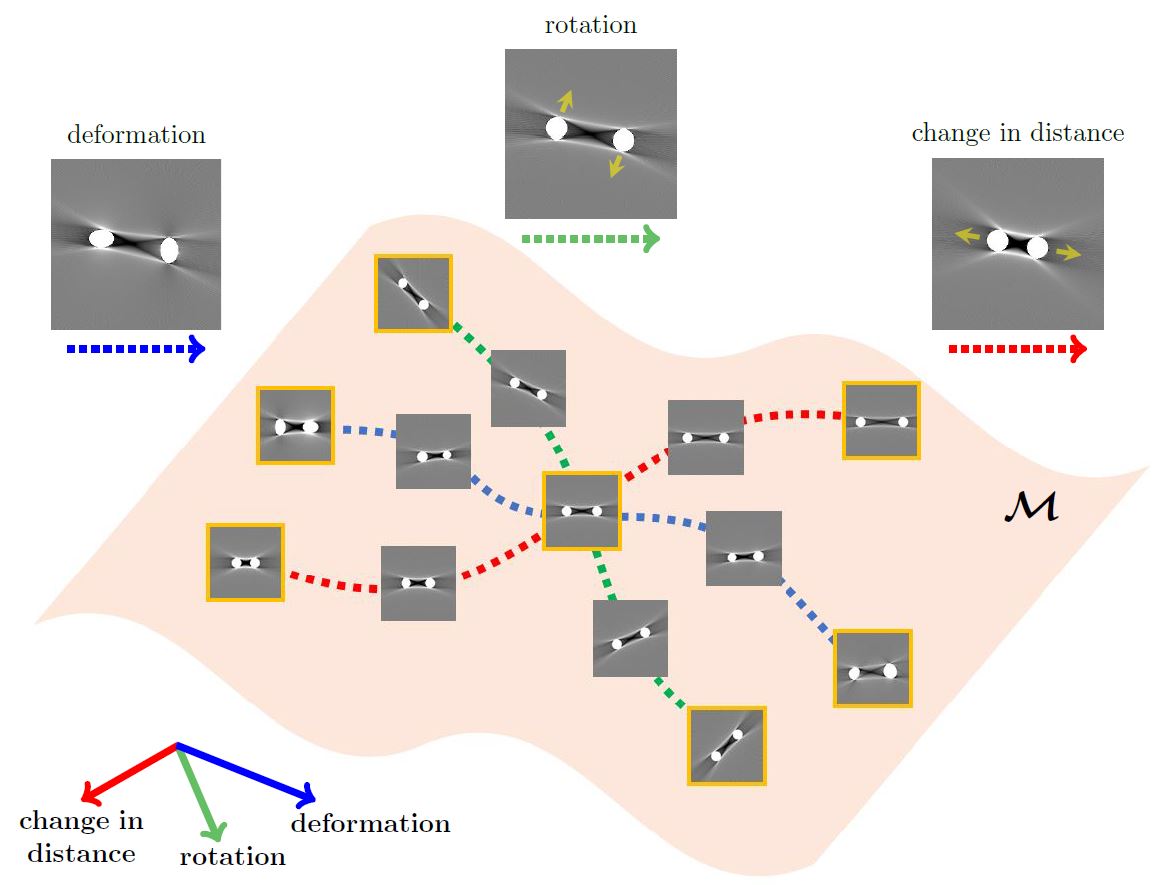}
\caption{Manifold of CT metal-artifact images. The trajectories in the embedding correspond roughly to smooth variations in metal artifacts with respect to rotation, deformation, and distance.}
\label{fig-manifold}
\end{figure*}

In general, deep learning methods require a considerable amount of training data. However, it is very difficult to obtain large amounts of real training data for MAR. We show the feasibility of applying a deep learning approach to MAR by using numerical simulations with real data. The proposed method was tested on real CT images of pelvises containing two simulated metallic (iron) inserts. In our simulations, $P_f$ and $P_{f^m}$ in \eref{P_f} and \eref{pfm} were generated with an X-ray tube voltage of $100$ kV with added Poisson noise. We used $\eta$ for the tungsten anode generated at a tube voltage of $100$ kVp \cite{Bushberg2002}. Other causes of metal artifacts, such as scattering and nonlinear partial volume effects were not considered. Here, $f_\CT$ and $f^m$ were acquired by applying FBP to $P_f$ and $P_{f^m}$, respectively. We took $f^m$ as the label. The input image $g^m$ was generated from \eref{gm}. Note that $\lambda$ and $c$ in \eref{gm} are associated with the $\eta$ and $\mu_m$ of metallic materials. Hence, we used fixed $\lambda$ and $c$ for all input $g^m$.
To train our network, we generated 690 CT images for two metallic objects in a symmetric position around the origin by changing their shape and position. More specifically, let $(\theta,r,(a_1,b_1,a_2,b_2))$ be a pair of transformation parameters for metallic objects to determine the shape and position. Here, $\theta$ denotes the rotation angle around the origin, $r$ is the distance from the origin to center of the metal, and $(a_1,b_1)$, $(a_2,b_2)$ denotes the pair of semi-major and semi-minor axes of the ellipse for each metal. We selected $\theta=(-43^\circ,-39^\circ,\cdots,45^\circ)$, $r=(7 cm,9 cm,\cdots,15 cm)$, and $(a_1,b_1,a_2,b_2) = (2,2,2,2)$, $(3,3,3,3)$, $(4,4,4,4)$, $(1,2,2,1)$, $(2,3,3,2)$, $(3,4,4,3)$, with an image size of 50$\times$50 cm
(see Fig. \ref{fig-manifold}). For the test images, we generated CT images for a simulated hip prosthesis, with geometry similar to the training set, obtained from pelvis CT images.

\subsection{Network setting}
The error \eref{error} was minimized using the RMSPropOptimizer \cite{Hinton2012} with a learning rate of 0.001, weight decay of 0.9, and mini-batch size of 20. We used 200 epochs to train the network. Training was implemented by Tensorflow \cite{Google} on a CPU (Intel(R) Core(TM) i7-6850K, 3.60GHz) and a GPU (NVIDIA GTX-1080, 8GB) system. It required approximately one hour to train our network.

\subsection{Numerical results for MAR}
\begin{figure}[ht]
\centering
\includegraphics[width=.5\textwidth]{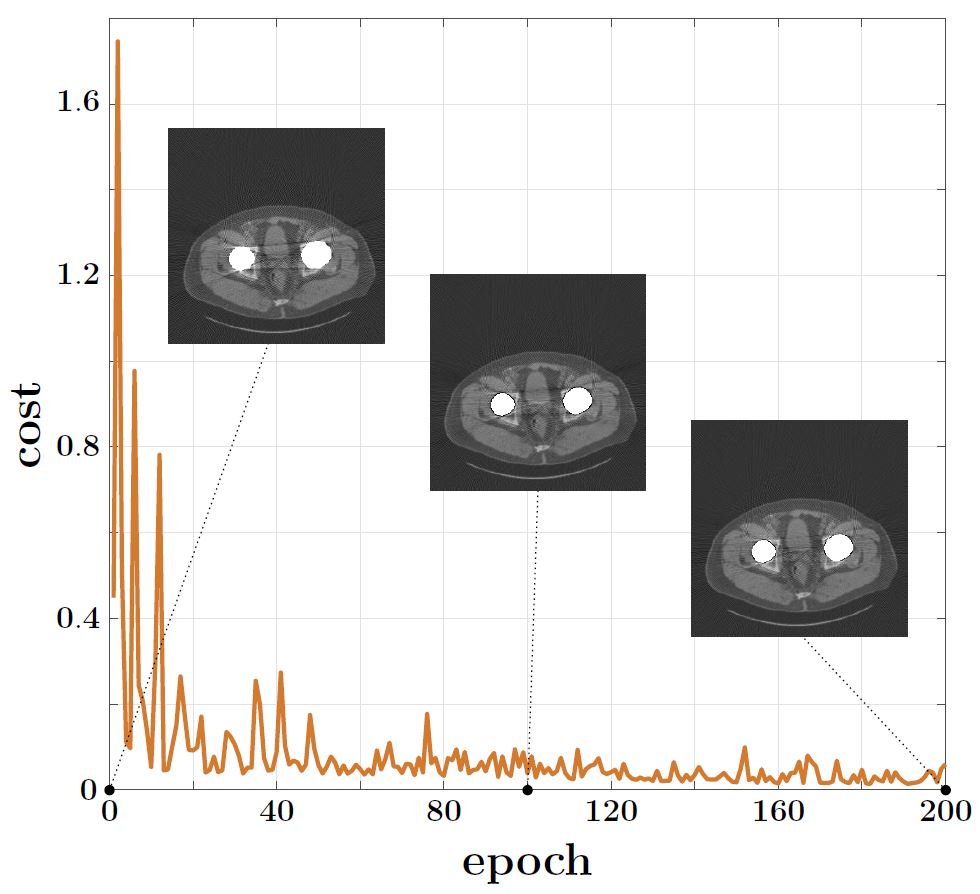}
\caption{Convergence plot for the $L^2$ error and test image results with respect to the number of epochs.}
\label{l2-loss}
\end{figure}

Fig. \ref{fig-manifold} shows a manifold $\mathcal M$ of CT metal-artifact images, which can be viewed as a set of points in high-dimensional space $\mathbb{R}^{256\times 256}$ with $256\times 256$ pixel CT images. Because linear combinations of two images on $\mathcal M$ do not resemble metal-induced beam-hardening artifacts, $\mathcal M$ is unlike Euclidean space. Our trained network performed well in terms of both the test set and training set. In other words, the regression manifold $\mathcal M$ effectively interpolated between CT metal artifact images in terms of their trajectories, as shown in Fig. \ref{fig-manifold}. Although we use a relatively small training data, the network produced reasonably accurate metal artifacts that were not in the training set. Fig. \ref{l2-loss} depicts the $L^2$ error and test image results with respect to the number of epochs. It shows that the $L^2$ error converged and that beam-hardening artifacts in test image were reduced as the number of epochs increased.

\begin{figure*}
\centering
\includegraphics[width=1\textwidth]{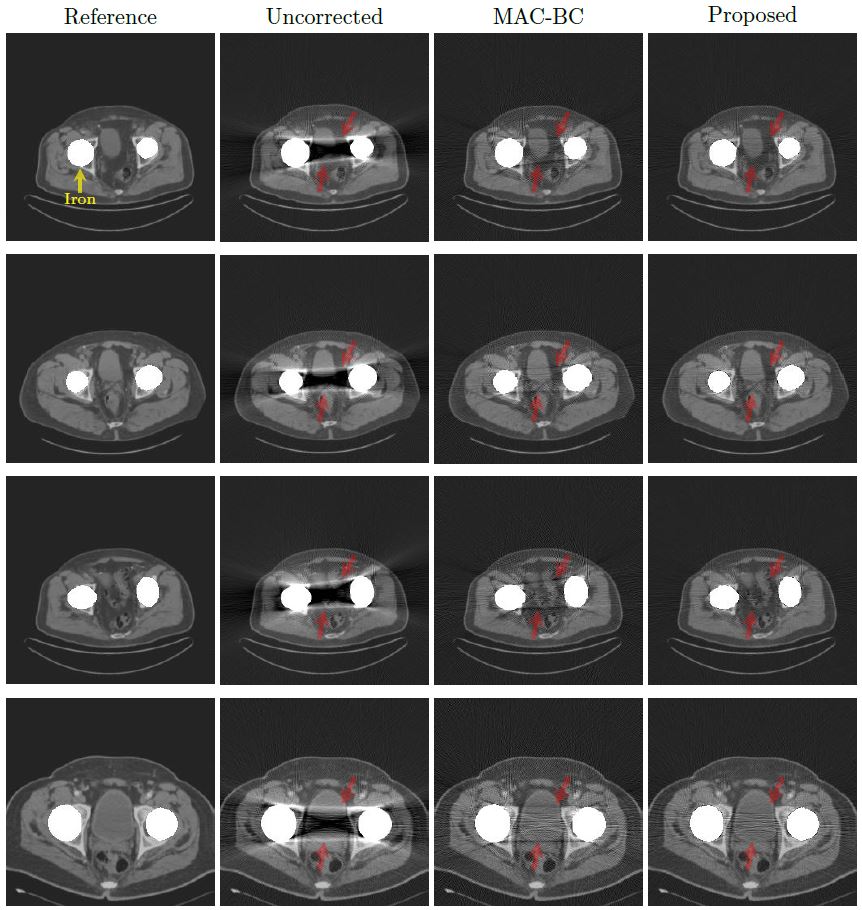}
\caption{Illustration of the numerical simulation results. The first and second columns show the reference and uncorrected image, respectively. The third and fourth columns represent the results with MAC-BC and the proposed method, respectively. The red arrows in the images mark the position of the artifacts. The results show that the proposed method reduce streaking artifacts more clearly than the MAC-BC method. (C=1000 HU/W=3000 HU.)}
\label{nm-result}
\end{figure*}

Fig. \ref{nm-result} shows the numerical simulation results for MAR. The first and second columns show the reference and $f_\CT$, respectively.  The third and fourth columns represent the correction results using MAC-BC and the proposed method, respectively. MAC-BC \cite{Park2016a} extracts beam-hardening artifacts from $f_\CT$ based on an analytic formula. Both correction methods show that the anatomical information corrupted by beam-hardening effects is clearly visible. However, MAC-BC still retained some streaking artifacts, owing to model's assumption of linear approximation for nonlinear variations (see Fig. \ref{nm-result}, third column). By comparison, the proposed method outperformed MAC-BC in terms of beam-hardening artifact reduction (see Fig. \ref{nm-result}, fourth column). In order to compare the quantitative errors in the corrected image ($f_{\text{cor}}$) from each method with those of the reference image ($f_{\text{ref}}$), we computed the normalized root mean square difference (NRMSD) \cite{Mehranian2013} on the outside of $ D_m$.
The NRMSD (\%) for each result is listed in Table \ref{table1}. The proposed method obtained the lowest NRMSD in all cases. This shows that the proposed method performs better at background restorations.
\begin{table}
\caption{NRMSD of the reconstructed image with the MAC-BC method and the proposed method}
\label{table1}
\centering
\begin{tabular}{c | c  c c c}
\hline\hline
&\quad Phantom&\quad Uncorrected&\quad MAC-BC&\quad Proposed\\
\hline
\multirow{4}{*}{NRMSD}
&\quad Pelvis 1&\quad 54.67&\quad 21.10&\quad 14.67\\
&\quad Pelvis 2&\quad 46.68&\quad 18.95&\quad 13.60\\
&\quad Pelvis 3&\quad 65.83&\quad 26.68&\quad 16.43\\
&\quad Pelvis 4&\quad 46.87&\quad 24.96&\quad 16.87\\
\hline
\end{tabular}
\end{table}

\section{Discussion and Conclusion}
We developed a deep learning MAR method to extract streaking and shadow artifacts from artifact-contaminated CT images arising from the presence of metallic objects. We demonstrated that the proposed method effectively provides nonlinear representations of shadows and streaking artifacts attributable to metallic objects. Ex-vivo features of metal artifacts were used for training rather than real in-vivo features, because there are no available ground-truth CT images for deep learning.

The proposed method works well with a relatively small number of training images, despite numerous variations to these artifacts in terms of the metal shape and placement.
We took advantage of the observation that metal-induced artifacts can be determined approximately from variations in metal geometry with respect to rotation, deformation, and distance. Hence, the collection of CT images of metal artifacts can be assumed to lie on a low-dimensional submanifold $\mathcal M$ embedded in high-dimensional space $R^{N}$, with the dimension $N$ denoting the number of pixels in the CT images. This observation allowed us to reduce the size of the training set significantly, with interpolation over the trajectory as shown in Fig. \ref{fig-manifold}.

However, there is room to further improve this method. The extraction procedure for metal artifacts could be modified by enhancing the forward model, which accurately represents real artifacts arising exclusively from metal inserts. Based on Observation \ref{prop1}, these real artifacts could be extracted directly from CT images, without any deep learning process.
Various experimental studies with patients will be required to explore the ability of this deep-learning-based approach, and we shall explore this in our future work.

\section*{Acknowledgements}
The authors (S. M. Lee, H. P. Kim, and J. K. Seo) was supported by Samsung Science $\&$ Technology Foundation (No. SSTF-BA1402-01). The first author (H. S. Park) was partially supported by the National Research Foundation of Korea(NRF) grant funded by the Korea government(Ministry of Science, ICT $\&$ Future Planning) (No. NRF-2016R1C1B2008098) and the National Institute for Mathematical Sciences (NIMS) grant funded by the Korean government (No. A21300000).


%

\bibliographystyle{IEEEtran}

\end{document}